\documentclass[10pt,superscriptaddress,twocolumn,amsmath,amssymb,aps,pra]{revtex4}
\usepackage{mathrsfs}
\usepackage{graphicx}
\usepackage{dcolumn}
\usepackage{bm}

\newcommand{\be}{\begin{equation}}
\newcommand{\ee}{\end{equation}}
\newcommand{\bea}{\begin{eqnarray}}
\newcommand{\eea}{\end{eqnarray}}

\begin{document}
\title{Anomalous Conductance of a strongly interacting Fermi Gas through a Quantum Point Contact}
\author{Boyang Liu}
\affiliation{Department of Physics and Center of Theoretical and
Computational Physics, The University of Hong Kong, Hong Kong,
China}

\author{Hui Zhai}
\affiliation{Institute for Advanced Study, Tsinghua University,
Beijing, 100084, China}

\author{Shizhong Zhang}
\affiliation{Department of Physics and Center of Theoretical and
Computational Physics, The University of Hong Kong, Hong Kong,
China}

\date{\today}
\begin{abstract}
In this work we study the particle conductance of a strongly interacting Fermi gas through a quantum point contact. With an atom-molecule two-channel model, we compute the contribution to particle conductance by both the fermionic atoms and the bosonic molecules using the Keldysh formalism. Focusing on the regime above the Fermi superfluid transition temperature, we find that the fermionic contribution to the conductance is reduced by interaction compared with the quantized value for the non-interacting case; while the bosonic contribution to the conductance exhibits a plateau with non-universal values that is larger than the quantized conductance. This feature is particularly profound at temperature close to the superfluid transition. We emphasize that the enhanced conductance arises because of the bosonic nature of closed channel molecules and the low-dimensionality of the quantum point contact.
\end{abstract}
 \maketitle
\section{Introduction}
The quantized conductance for transport through a quantum point contact (QPC) is one of the most prominent phenomena in mesoscopic physics~\cite{Wees,Wharam}. The quantization in units of $2e^2/h$ is well understood in terms of the
Landauer-B\"{u}ttiker formalism~\cite{Landauer, Buttiker} for non-interacting electrons, where $e$ is the electric charge and $h$ is the Planck's constant.  Recently, experimental advances in cold atom systems have made it possible to engineer mesoscopic tunnel junctions between two reservoirs of degenerate quantum gases. The advantage of this novel system is that one can tune the inter-atomic interaction to study the interaction and correlation effects in the mesoscopic transport phenomena in a highly controllable manner.  In 2015, the quantized conductance of non-interacting neutral Fermi atoms was observed by the ETH group where the conductance is found to be quantized in units of $1/h$~\cite{Krinner1}. Subsequently, the transport properties of strongly interacting Fermi gases through QPCs were also investigated experimentally for both the superfluid state~\cite{Husmann} and the normal state~\cite{Krinner2}.

Of particular interest is the anomalous conductance discovered in the normal state of a strongly interacting Fermi gas through a QPC~\cite{Krinner2}. It is found that the height of the conductance plateau can be enhanced to a non-universal value, which can be several times of $1/h$, as the interaction strength increases. This anomalous conductance is qualitatively different from the conductance anomalies observed in solid state QPC, where it is usually {\em reduced} due to electron correlations. In this paper, we put forward one possible explanation for the enhanced conductance observed experimentally. We show that the presence of molecular state, or preformed pairs, in the strongly interacting Fermi gas can contribute to a larger conductance due to the Bose statistics~\cite{Stringari}, even above the superfluid transition temperature $T_\text{c}$.

In this work the bulk properties of the strongly interacting Fermi gas in two reservoirs are described by an atom-molecule coupled two-channel model, which takes explicitly into account both fermionic and bosonic degrees of freedoms \cite{Holland,Timmermans,Ohashi}. We focus on the normal state transport through a QPC at unitarity and adopt the Nozi\`{e}res and Schmitt-Rink (NSR) scheme \cite{Ohashi,NSR} to calculate the spectral functions of both bosonic and fermionic components in the two reservoirs. Then we employ a tunneling Hamiltonian to describe the particle transport through a QPC, and calculate the current contributions from both the bosonic and fermionic components using the Keldysh formalism~\cite{Keldysh,Kamenev}. We find that the conductance still processes quantized plateaux. The conductance of the fermionic component is reduced due to the strong interaction effect, while that by bosons is enhanced, and this enhancement becomes stronger as one approaches the critical temperature.

\section{The model}
\begin{figure}[ht]
\begin{center}
  \includegraphics[width=8cm]{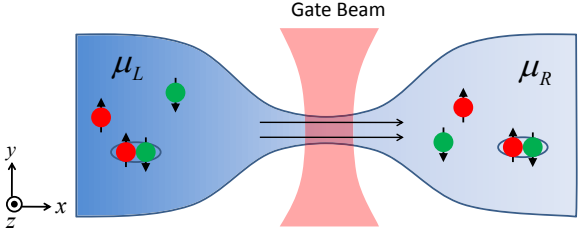}
  \caption{The geometry of the experiment setup~\cite{Krinner2}. Two reservoirs are connected by a quasi-one-dimensional channel, through which the fermionic spin-$1/2$ atoms and bosonic molecules can tunnel from one side to another. A gate beam applied on the central regime tunes the particle density in the quasi-one-dimensional channel.}
  \label{fig:exp}
  \end{center}
 \end{figure}

In the realistic experiment setup the cigar-shaped trap is split into two reservoirs using the repulsive potential of a TEM01-like mode of a laser. A two-dimensional channel is formed between the two reservoirs initially. Then, at the center of the two-dimensional region a QPC is created by imaging a split gate structure using high-resolution lithography \cite{Krinner1,Husmann,Krinner2}. The particles will be transferred from the three-dimensional reservoirs to a two-dimensional region first, and then enters the one-dimensional QPC. To model the tunneling process, we simplify the experimental structure as shown in Fig.~\ref{fig:exp}: Two three-dimensional reservoirs are connected directly by a one-dimensional QPC. The transverse trapping frequencies $\omega_{y,z}$ of the QPC in the $y$ and $z$ directions are much larger than the thermal energy of the system, and as a result, the tunneling channel can be regarded as one-dimensional. An additional laser beam is shone on the QPC along the $y$ direction to create an attractive gate potential $V_g$ which can tune the particle density in this region.

The total Hamiltonian of the system, consisting of two reservoirs and the QPC, can be written as (setting $\hbar=1$)
\be
\hat{H}=\hat{H}_L+\hat{H}_R+\hat{H}_T,
\label{eq:Ham}
\ee
where $\hat{H}_L (\hat{H}_R)$ describes the left (right) reservoir and is given by
\begin{align}
\hat{H}_j =\int d^3{\bf r}\Big\{&\sum_{\sigma}\hat{\psi}^\dag_{j\sigma}({\bf r})\big(-\frac{\nabla^2}{2m}-\mu_j\big)\hat{\psi}_{j\sigma}({\bf r})\cr&+\hat{\phi}^\dag_{j}({\bf r})\big(-\frac{\nabla^2}{4m}+2\nu-2\mu_j\big)\hat{\phi}_{j}({\bf r})\nonumber\\
&+g\Big(\hat{\psi}^\dag_{j\uparrow}({\bf r})\hat{\psi}^\dag_{j\downarrow}({\bf r})\hat{\phi}_{j}({\bf r})+{\rm h.c.}\Big)\Big\}
\label{eq:twochannel}
\end{align}
The operator $\hat{\psi}_{j\sigma}({\bf r})$ describes the creation of a fermion atom with spin $\sigma=\uparrow,\downarrow$ in the $j$-th reservoir with $j=L,R$.  $m$ is the mass of fermions. Similarly, $\hat{\phi}_{j}({\bf k})$ creates a diatomic molecule of mass $2m$ in the $j$-th reservoir. $2\nu$ is the bare detuning of the molecular state with respect to scattering continuum and $g$ is the bare coupling constant between the atoms and the molecules. These two parameters can be related to the $s$-wave scattering length $a_s$ and the effective range $r_0$:
\begin{align}
\frac{2\nu}{g^2} &=-\frac{m}{4\pi a_s}+\int\frac{d^3{\bf k}}{(2\pi)^3}\frac{1}{2\epsilon_{\bf k}},\\
\frac{1}{g^2} &=-\frac{r_0 m^2}{8\pi},
\end{align}
where $\epsilon_{\bf k}={\bf k}^2/2m$.

Finally, $\hat{H}_T$ describes the tunneling between two reservoirs through the QPC. Within the QPC, the transverse confinement leads to the transport channels with energies given by $\epsilon_\perp(n_y,n_z)\equiv (\frac{1}{2}+n_y)\omega_y+(\frac{1}{2}+n_z)\omega_z+V_g$. We define an effective gate potential as $\bar V_g=V_g+\frac{1}{2} \omega_y+\frac{1}{2} \omega_z$ and assume $\omega_z\gg\omega_y$, then the eigen-energy of the several lowest transport channels would be $n_y \omega_y+\bar V_g$, and they are non-degenerate. The potential along the transport direction ($\hat{x}$-direction in our case, see Fig.\ref{fig:exp}) can be modeled as a saddle point potential with $V(x)=-\frac{1}{2}m\omega_x^2 x^2$ [see Figure 1(c) of Ref.~\cite{Krinner2}], neglecting the modification of potential to the entry and exit of the QPC. In our case, $\omega_x\ll \omega_{y,z}$ and for this particular case, it is known that transmission matrix element through the $n$-th channel of the QPC is energy dependent, $\mathcal{T}^n(E)=(1+\exp[-2\pi(E-n\omega_y-\bar{V}_g)/\omega_x])^{-1}$~\cite{Buttiker90}. We assume that only particle with momentum along the $\hat{x}$-direction can pass through the QPC, and away from the QPC, their energies are given by $\epsilon(k_{j,x})=k_{j,x}^2/2m$, $j=L,R$. As a result, we can make the following simplification for the tunneling matrix elements when $\omega_x\ll\omega_{y,z}$
\begin{align}
\mathcal{T}^n_{F}(k_{L,x},k_{R,x}) &=\mathcal{T}_{F}\prod_{j=L,R}\Theta\left[\epsilon(k_{j,x})-n \omega_y-\bar{V}_g\right],\\
\mathcal{T}^n_{B}(k_{L,x},k_{R,x}) &=\mathcal{T}_{B}\prod_{j=L,R}\Theta\left[\frac{\epsilon(k_{j,x})}{2}-n \omega_y-\bar{V}_g\right].
\end{align}
The constants $\mathcal{T}_{F,B}$ will be related to the transparency of the QPC later. The tunneling matrix elements above indicate that only particles with energy $\epsilon(k_{j,x})>n \omega_y+\bar{V}_g$ can enter the $n$-th channel of the QPC and will come out from the same channel. Namely, we assume that there is no inter-channel scattering within the QPC. With this simplification, we can write the tunneling Hamiltonian in the explicit form [now written in Heisenberg representation. For more detail, see Appendix A],
\begin{align}\nonumber
&\hat{H}_T=\frac{1}{A^2}\sum_{n}\int\frac{d\omega}{2\pi}\frac{dk_{L,x}}{2\pi }\frac{dk_{R,x}}{2\pi}\{\\\nonumber
&\mathcal{T}_F^n(k_{L,x},k_{R,x})\sum_\sigma\hat{ \psi}^\dag_{L\sigma}(\omega, k_{L,x})\hat{\psi}_{R\sigma}(\omega+\Delta\mu,k_{R,x})+\\
&\mathcal{T}^n_B(k_{L,x},k_{R,x})\hat{\phi}^\dag_{L}(\omega,k_{L,x})\hat{\phi}_{R}(\omega+2\Delta\mu, k_{R,x})+{\rm h.c.}\}.
\label{eq:tunnel}
\end{align}
Here $A$ is the cross-section area in the $yz$ plane of the QPC.  $\Delta\mu=\mu_L-\mu_R$ are the chemical potential bias, where $\mu_{L,R}$ is the chemical potentials for fermions of the left and right reservoirs.

\section{The Conductance Formalism}
In the presence of a chemical potential bias $\Delta \mu$, the total (atomic) current is a sum of both the fermionic and bosonic parts as $I(t)=I_F(t)+I_B(t)$, where
\begin{align}
I_F &\equiv\frac{1}{2}\left\langle\frac{\partial}{\partial t}(N_{F,R}-N_{F,L})\right\rangle,\\
I_B &\equiv\left\langle\frac{\partial}{\partial t}(N_{B,R}-N_{B,L})\right\rangle,
\end{align}
with $N_{F,j}\equiv \sum_{{\bf k}\sigma}\hat{\psi}^\dag_{j\sigma}({\bf k})\hat{\psi}_{j\sigma}({\bf k})$ and $N_{B,j}=\sum_{{\bf k}}\hat{\phi}^\dag_{j}({\bf k})\hat{\phi}_{j}({\bf k})$ are the number operators of fermions and bosons in the two reservoirs, respectively. In the above expressions the averages $\langle\cdot\cdot\cdot\rangle$ is taken over a time-evolving many-body state, which we implement using the Keldysh formalism~\cite{Husmann,Keldysh,Kamenev}. To second order in $\mathcal T_{F(B)}$, one obtains the expressions for the currents $I_{F,B}(t)$ in terms of spectral functions $A_j^{F,B}(\omega,{\bf k})$ for fermions and bosons in the reservoirs (see Appendix B)
\begin{align}
&I_{F}(t)=2\alpha_{F}\epsilon_F^R\int \frac{d\omega}{2\pi}\frac{d\epsilon_L}{\sqrt{\epsilon_L}}\frac{d\epsilon_R}{\sqrt{\epsilon_R}} \Theta(\epsilon_L-n \omega_y-\bar V_g)\cr
&\Theta(\epsilon_R-n \omega_y-\bar V_g)A^{F}_L(\omega,\sqrt{2m\epsilon_L}) A^{F}_R(\omega+\Delta\mu,\sqrt{2m\epsilon_R})\cr&\big[n_{F}(\omega)-n_{F}(\omega+\Delta\mu)\big],
\label{eq:one-dimensionalcurrent}
\end{align}
where $\epsilon^R_F$ is the Fermi energy of the right reservoir. In this work we will use $\epsilon^R_F$ and the Fermi momentum $k_F^R$ of the right reservoir as the energy and the momentum units. The expression for $I_\text{B}$ is similar except that $m$, $\alpha_{F}$, $A^{F}_j$ and $n_{F}$ are now replaced by $2m$, $\alpha_{B}$, $A^{B}_j$ and $n_{B}$. Furthermore for $I_B(t)$, $\Delta\mu_F\equiv \Delta\mu$ should be replaced by $\Delta\mu_B\equiv 2\Delta\mu$.  The so-called transparency is defined as $\alpha_{F(B)}=16\pi^2|\mathcal {T}_{F(B)}|^2D^2_{F(B)}(\epsilon^R_F)/A^2$, where $D_F(\epsilon)=\sqrt{2m}/(4\pi\sqrt\epsilon)$ and $D_B(\epsilon)=\sqrt{4m}/(4\pi\sqrt\epsilon)$ are the one-dimensional density of state for fermions and bosons, respectively. In this work, we assume perfectly transparent junction and set both $\alpha_{F,B}=1$ for simplicity. The atomic conductances are then calculated by $\sigma=\sigma_F+\sigma_B$ with $\sigma_{F,B}=I_{F,B}/\Delta\mu$.

For non-interacting fermions, the spectral functions are $\delta$-functions: $A^{F}_L(\omega,\sqrt{2m\epsilon_L})=\delta(\omega-\epsilon_L+\mu_L)$ and $A^{F}_R(\omega+\Delta\mu,\sqrt{2m\epsilon_R})=\delta(\omega+\Delta\mu-\epsilon_R+\mu_R)$. Then the current in Eq. (\ref{eq:one-dimensionalcurrent}) reduces to the standard Landauer-B\"{u}ttiker formula~\cite{Landauer, Buttiker,Kamenev} as $I_F=\frac{1}{\pi}\int d\epsilon\Theta(\epsilon-n\omega_y-\bar V_g)[n_{F}(\epsilon-\mu_L)-n_{F}(\epsilon-\mu_R)]$, which exhbits quantized conductance plateaux as $\bar{V}_g$ changes. For the interacting cases, the spectral functions of fermions and bosons are given by
\begin{align}
A^F_j(\omega,{\bf k}) &=-\frac{1}{\pi}{\rm Im}\frac{1}{\omega-\epsilon_{\bf k}+\mu_j-\Sigma^F_j(\omega,{\bf k})},\\
A^B_j(\omega,{\bf k}) &=-\frac{1}{\pi}{\rm Im}\frac{1}{\omega-\epsilon_{\bf k}/2-2\nu_0+2\mu_j-\Sigma^B_j(\omega,{\bf k})}.
\end{align}
Here $\Sigma^{F,B}_j(\omega,{\bf k})$ is the self-energies of the bosonic and fermionic components in the $j$-th reservoir and will be calculated within NSR scheme in the following.

\section{The spectral weight functions of fermions and bosons in NSR scheme}
Within the NSR scheme~\cite{Ohashi,NSR}, the number density in $j$-th reservoir is given by
\begin{align}\nonumber
n_j =& 2\int\frac{d^3 {\bf k}}{(2\pi)^3}n_F(\epsilon_{\bf k}-\mu_j)\\
&+\int\frac{d^3 {\bf k}}{(2\pi)^3}\frac{d\omega}{\pi}n_B(\omega)\frac{\partial}{\partial\mu_j}\delta_j(\omega,{\bf k}),
\end{align}
where $n_{F(B)}(\omega)=(\exp[\beta \omega]\pm1)^{-1}$ is the Fermi (Bose) distribution function. The phase shift is defined by
\be
\delta_j(\omega,{\bf k})=-\mbox{arg}\Big[\frac{-\omega+\frac{\epsilon_{\bf k}}{2}+2\nu_0-2\mu_j}{g^2}-\Pi_j(\omega,{\bf k})\Big],
\ee
where arg means taking the phase of the expression in the bracket. Here $\nu_0=-mg^2/(8\pi a_s)$ and the polarization $\Pi_j$ is given by
\be
\Pi_j=\int\frac{d^3{\bf p}}{(2\pi)^3}\Big[\frac{1-n_F(\epsilon_{\bf p}-\mu_j)-n_F(\epsilon_{{\bf p}-{\bf k}}-\mu_j)}{-\omega+\epsilon_{\bf p}+\epsilon_{{\bf p}-{\bf k}}-2\mu_j}-\frac{1}{2\epsilon_{\bf p}}\Big].\label{eq:polarization}
\ee
We consider transport when both reservoirs are in the normal state above the superfluid transition temperature $T_c$, which in turn is given by  the Thouless criterion
\begin{align}
-\frac{2\nu_0-2\mu_j}{g^2}+\int\frac{d^3k}{(2\pi)^3}\left[\frac{1-2n_F(\epsilon_{\bf k}-\mu_j)}{2\epsilon_{\bf k}-2\mu_j}-\frac{1}{2\epsilon_{\bf k}}\right]=0.
\end{align}

\begin{figure}[h]
\begin{center}
  \includegraphics[width=8cm]{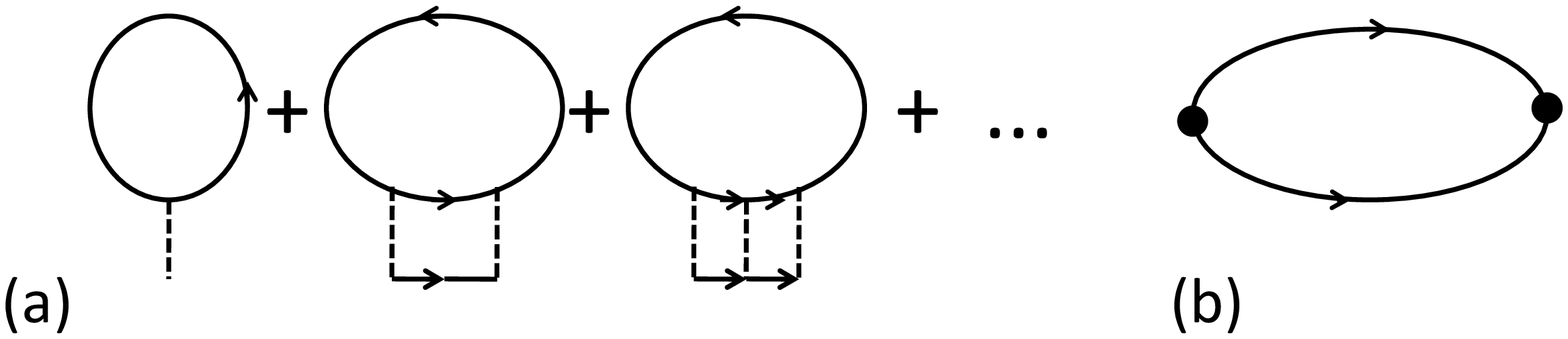}
  \caption{ The Feynman diagrams for self-energies of (a) fermions and (b) bosons. Dashed lines and solid lines denote the boson and fermion propagators, respectively. }
    \label{fig:SE}
  \end{center}
 \end{figure}
The self-energies of fermions and bosons can be calculated from the Feynman diagrams in Fig. \ref{fig:SE}. The fermion loop in Fig. \ref{fig:SE} (b) is the polarization operator $\Pi_j(\omega, {\bf k})$ in Eq. (\ref{eq:polarization}).
\begin{align}
&\Sigma_j^B(\omega+i0^+,{\bf k}) \\\nonumber
=&-g^2\Pi_j(\omega+i0^+,{\bf k})\\\nonumber
=&g^2\int\frac{d^3{\bf p}}{(2\pi)^3}[\frac{1-n_F(\epsilon_{\bf
p}-\mu_i)-n_F(\epsilon_{{\bf p}-{\bf k}}-\mu_i)}{\omega+i0^+-\epsilon_{\bf
p}-\epsilon_{{\bf p}-{\bf k}}+2\mu_i-i0^+}+\frac{1}{2\epsilon_{\bf p}}].
\end{align}

The self-energy of fermion can be calculated by summing the diagrams in Fig. \ref{fig:SE} (a) as
\bea
&&\Sigma_j^F(i\omega_n,{\bf k})=\frac{1}{\beta}\sum_{\omega_m}\int\frac{d^3{\bf p}}{(2\pi)^3}\frac{1}{-i(\omega_m-\omega_n)+\epsilon_{\bf p-k}-\mu_j}\cr&&\Big\{-U(i\omega_m,{\bf p})-U(i\omega_m,{\bf p})\Pi(i\omega_m,{\bf p}) U(i\omega_m,{\bf p})\cr&&-U(i\omega_m,{\bf p})\Pi(i\omega_m,{\bf p}) U(i\omega_m,{\bf p})\Pi(i\omega_m,{\bf p}) U(i\omega_m,{\bf p})+...\Big\}\cr&&=\frac{1}{\beta}\sum_{\omega_m}\int\frac{d^3{\bf p}}{(2\pi)^3}\frac{1}{-1/U(i\omega_m,{\bf p})+\Pi(\omega_m,{\bf p})}\cr&&\frac{1}{-i(\omega_m-\omega_n)+\epsilon_{\bf p-k}-\mu_j},\eea
where $U(i\omega_m,{\bf p})={g^2}({-i\omega_m+p^2/4m+2\nu_0-2\mu_j})^{-1}$.  After the analytical continuation we obtain the fermion self-energy as
\begin{align}
&\Sigma_j^F(\omega+i0^+,{\bf k})=\frac{1}{\beta}\sum_{\omega_m}\int\frac{d^3{\bf p}}{(2\pi)^3}\times\cr
&\frac{1}{\chi(\omega_m,{\bf p})[\omega+i0^+-i\omega_m+({\bf p}-{\bf k})^2/2m-\mu_j]},
\end{align}
where we have defined
\bea\chi(\omega_m,{\bf p})=-1/U(i\omega_m,{\bf p})+\Pi(\omega_m,{\bf p}).\eea

\begin{figure}[t]
\begin{center}
  \includegraphics[width=8cm]{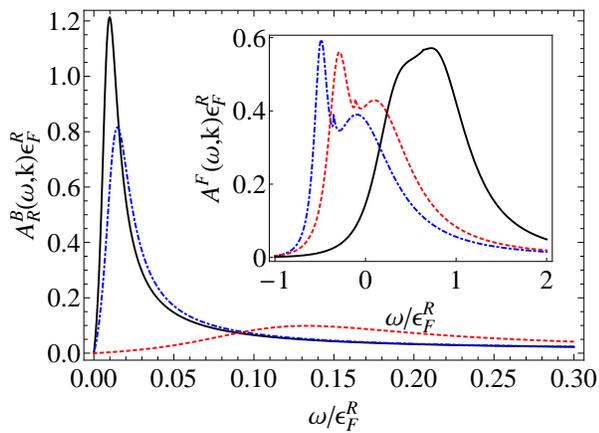}
  \caption{The spectral weight function of bosons $A^B_R(\omega,k)$. The black solid, blue dot-dashed and red-dashed lines are for three different momenta $k/k_F^R=0.01, 0.1, {\rm and}~ 0.5$. The inset is the spectral weight function of fermions $A^F_R(\omega,k)$. The blue dot-dashed, red-dashed and black solid lines are for momenta $k/k_F^R=0.1, 0.5, {\rm and}~ 1$. The scattering length is set at resonance when $1/(a_sk_F^R)=0$ and the effective range $k_F^Rr_0=-0.1$. The temperature is above the superfluid transition temperature with $T/T_c^R=1.1$.}
  \label{spectral}
  \end{center}
 \end{figure}

The spectral function for bosons is shown in Fig. \ref{spectral} for three different momenta. It is worth noting that the accumulation of spectral weight at low energy when $k$ is small. It is this low energy spectral weight that contributes an enhanced bosonic conductance through the QPC. On the other hand, the Fermi spectral function within the NSR calculation shows broad peaks. For $k=k_F$ (black line in inset of Fig.\ref{spectral}), the width of the spectral function is comparable to Fermi energy, indicating that Landau quasi-particle is not well defined within NSR. In addition, the peak position occurs away from $\omega=0$, consistent with psuedo-gap behavior.

\begin{figure}[t]
\begin{center}
  \includegraphics[width=9cm]{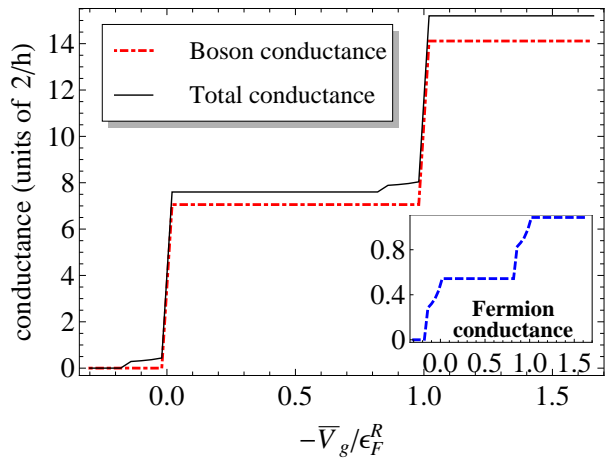}
  \caption{The conductances contributed by bosons and fermions (inset), and the total conductance as functions of the effective gate potential $\bar V_g$. Both boson and fermion conductances exhibit plateaux. The conductance is enhanced for bosons while suppressed for fermions. We focus on the resonance with $1/(a_sk_F^R)=0$ and the effective range $k_F^Rr_0=-0.1$, and $T/T_c^R=1.1$. }
  \label{fig:condvg}
  \end{center}
 \end{figure}

\section{The Conductance Results}
In Fig.\ref{fig:condvg} we plot the conductances contributed by both the bosons and fermions, as well as the total conductance as functions of the effective gate potential $\bar V_g$. The first feature to be noted is that the conductances for both bosons and fermions still exhibit plateaux. However, the height of fermion conductance plateaux is reduced to a smaller values than the non-interacting case due to strong interaction which is consistent with a non-Fermi liquid behavior at unitarity. 

On the other hand, the conductance contributed by bosons is much higher. There are two effects leading to this large bosonic conductance: (i) The boson spectral weight distribution and the Bose statistics. In Fig. \ref{spectral}, we show that the spectral function of the bosonic molecule is sharply peaked at small momentum at low energy in the strongly interacting regime. Meanwhile, the weight of the Bose distribution function $n_B(\omega)=(\exp[\beta\omega]-1)^{-1}$ increases towards the small $\omega$ limit, and this leads to a large value of boson current in Eq. (\ref{eq:one-dimensionalcurrent}). (ii) The low dimensional structure of the QPC. If the QPC is just a point in three-dimension, the integration over the momenta is equal to an average over a three-dimensional density of state, which vanishes as the energy approaches zero, which would have canceled the enhancement discussed in (i). However, in the one-dimensional tunneling channel we are discussing here, the density of state at low-energy is finite and even diverges when approaching zero, which guarantees non-vanishing contribution and the existence of the enhancement effect, consistent with conclusion reached in Ref. \cite{Stringari}.

\begin{figure}[t]
\begin{center}
  \includegraphics[width=8cm]{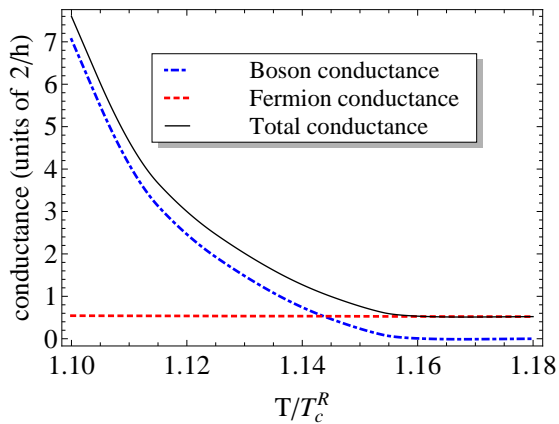}
  \caption{(Color online) The conductances of bosons and fermions, as well as the total conductance as functions of the temperature. The scattering length $1/a_sk_F^R=0$ and the effective range $k_F^Rr_0=-0.1$. Here we the conductance is calculated at $\bar V_g/\epsilon_F^R=-0.5$.}
  \label{fig:condT}
  \end{center}
 \end{figure}

In Fig. \ref{fig:condT}, we show the variation of the conductances as a function of temperature $T>T_{c}$. We observe that the conductance of bosons drops to zero very fast as one increases the temperature. The effect of anomalous conductance is thus most pronounced when temperature is close to $T_\text{c}$.

\section{Discussions}
Throughout this work, the conductance through a QPC is presented for a ``narrow" Feshbach resonance, where the effective range is taken as $k^R_Fr_0=-0.1$. We find that the contribution from the bosonic molecules can give rise to an anomalous large conductance in a strongly interacting Fermi gas. For a broad resonance, our calculation shows that the closed channel molecule fraction is about $\int d\omega A^B(\omega, k)\simeq10^{-4}$ for typical value of $k$ for a broad resonance of $^6$Li where $k_F^R r_0=-0.00016$  \cite{Partridge}. This small fraction significantly reduces the conductance contributed by bosons. Nevertheless, this does not mean the contribution from the bosonic degree of freedom is not important for a broad Feshbach resonance, for which the bosonic degrees of freedom exist primarily as the fluctuating Cooper pairs of open channel character.

When finishing this paper, two other theoretical works also propose explanations for this anomalous conductance~\cite{Glazman,Uchino}. What is common of these three papers is that they all emphasize the role of bosonic degree of freedoms. Ref.~\cite{Glazman} assumes that in the QPC, the strong confinement renormalizes the interaction such that pairing occurs in the channel, for which they emphasize the role of multichannel Andreev reflections. While both Ref.~\cite{Uchino} and our work do not assume pairing in the channel but focus on the effects of pairing fluctuation in reservoirs. Ref.~\cite{Uchino} uses a single channel model while our work uses a two-channel model. In their calculation they also find that the fluctuation effect suppresses the transport of fermionic particles, while the contribution from the fluctuating pairs is enhanced~\cite{Uchino}. This is consistent with our conclusion that the fermionic contribution is suppressed while the bosonic contribution is enhanced. Hence, Ref.~\cite{Uchino} and our work can be regarded as complementary to each other.

\section{Acknowledgements}
We would like to thank Thierry Giamarchi, Shun Uchino and Haizhou Lu for discussions. B.Y. and S.Z. are supported by Hong Kong Research Grants Council, General Research Fund, HKU 17306414 and the Croucher Foundation under the Croucher Innovation Award. H.Z. is supported by MOST under Grant No. 2016YFA0301600, NSFC Grant No. 11325418 and Tsinghua University Initiative Scientific Research Program.

\appendix
\section{The tunneling Hamiltonian}
To discuss the construction of the tunneling Hamiltonian, we start with the real space tunneling Hamiltonian at point ${\bf r}=0$.
\begin{equation}
H_T=\sum_{n,\sigma}[\mathcal{T}^n_F\bar\psi_{L\sigma}(0)\psi_{R\sigma}(0)+\mathcal{T}^n_B\phi^\ast_{L}(0)\phi_{R}(0)]+h.c.
\end{equation}
In order to investigate $H_T$ in the Keldysh framework, we need to derive the time evolution of $H_T$. Due to the chemical potential difference in two reservoirs, we use the single particle Hamiltonian $H+\sum_j(\mu_j N_{F,j}+2\mu_jN_{B,j})$ to construct the time evolution operator as
\begin{equation}
U(t)=e^{i[H+\sum_j(\mu_j N_{F,j}+2\mu_jN_{B,j})]t}
\end{equation}
In this case, the energies in both reservoirs are measured on the  absolute scale. Then the time evolution of $H_T$ is given by
\begin{align}
&H_T(t) =U(t)H_TU^{-1}(t)\\
=&\sum_n[\mathcal{T}^n_Fe^{-it\Delta\mu}\sum_\sigma\bar\psi_{L\sigma}(t,0)\psi_{R\sigma}(t,0)\\\nonumber
&+\mathcal{T}^n_Be^{-2it\Delta\mu}\phi^\ast_{L}(t,0)\phi_{R}(t,0)]+h.c.,
\end{align}
where $\psi_{j\sigma}(t,0)=e^{iHt}\psi_{j\sigma}(0)e^{-iHt}$ and $\phi_{j}(t,0)=e^{iHt}\phi_{j}(0)e^{-iHt}$. In momentum space the tunneling Hamiltonian is written as
\begin{align}\nonumber
&\hat{H}_T=\sum_{n}\int\frac{d\omega}{2\pi}\frac{d{\bf k}_L}{(2\pi)^3 }\frac{d{\bf k}_{R}}{(2\pi)^3}\{\\\nonumber
&\mathcal{T}_F^n({\bf k}_{L},{\bf k}_{R})\sum_\sigma\hat{ \psi}^\dag_{L\sigma}(\omega, {\bf k}_{L})\hat{\psi}_{R\sigma}(\omega+\Delta\mu,{\bf k}_{R})+\\
&\mathcal{T}^n_B({\bf k}_{L},{\bf k}_{R})\hat{\phi}^\dag_{L}(\omega,{\bf k}_{L})\hat{\phi}_{R}(\omega+2\Delta\mu, {\bf k}_{R})+{\rm h.c.}\}.
\end{align}
In the momentum space representation above, it's clearly shown that a fermion with energy $\omega$ in the left reservoir will tunnel through the QPC and ends up as a fermion with energy $\omega+\Delta\mu$ in the right reservoir. Similar considerations applies for bosons. Furthermore, in our model we assume only particle moving along $\hat x$-direction can pass through the QPC. The three-dimensional momentum integration will be reduced to one-dimensional integration by $\int\frac{d{\bf k}}{(2\pi)^3}\rightarrow\frac{1}{A}\int\frac{dk_x}{2\pi}$, where $A$ is the cross-section area of the QPC. Then the tunneling Hamiltonian can be written as
\begin{align}\nonumber
&\hat{H}_T=\frac{1}{A^2}\sum_{n}\int\frac{d\omega}{2\pi}\frac{dk_{L,x}}{2\pi }\frac{dk_{R,x}}{2\pi}\{\\\nonumber
&\mathcal{T}_F^n(k_{L,x},k_{R,x})\sum_\sigma\hat{ \psi}^\dag_{L\sigma}(\omega, k_{L,x})\hat{\psi}_{R\sigma}(\omega+\Delta\mu,k_{R,x})+\\
&\mathcal{T}^n_B(k_{L,x},k_{R,x})\hat{\phi}^\dag_{L}(\omega,k_{L,x})\hat{\phi}_{R}(\omega+2\Delta\mu, k_{R,x})+{\rm h.c.}\}.
\end{align}

\section{The Keldysh formalism of the tunneling Hamiltonian}
To study the particle transport through the QPC, we follow the Keldysh formalism on a closed time contour~\cite{Kamenev}. The action of the left and right reservoir can be written as
\begin{align} \nonumber
S_i = &\int d{\bf x} dt\Big\{\sum_\sigma\bar\Psi_{j\sigma} [G_{Fj}]^{-1}\Psi_{j\sigma}+\Phi^\ast_j [G_{Bj}]^{-1}\Phi_j\\\nonumber
&+g\big[\phi^{\rm cl}_j(\bar\psi_{j1\uparrow}\bar\psi_{j2\downarrow}+\bar\psi_{j2\uparrow}\bar\psi_{j1\downarrow})\\
&+\phi^{\rm q}_j(\bar\psi_{j1\uparrow}\bar\psi_{j1\downarrow}+\bar\psi_{j2\uparrow}\bar\psi_{j2\downarrow})+h.c.\big]\Big\},
\end{align}
where the bosonic and fermionic coherent fields are given by
\begin{align}
\bar\Psi_{j\sigma} &=\left(\begin{array}{cc}\bar\psi_{j1\sigma}&\bar\psi_{j2\sigma}\end{array}\right)\\
\Psi_{j\sigma} &=\left(\begin{array}{c}\psi_{j1\sigma}\\ \psi_{j2\sigma}\end{array}\right)\\
\Phi_{j}^\ast &=\left(\begin{array}{cc}{\phi^{\rm cl}_{j}}^\ast& {\phi_{j}^{\rm q}}^\ast\end{array}\right)\\
\Phi_{j} &=\left(\begin{array}{c}\phi^{\rm cl}_{j}\\ \phi^{\rm q}_{j}\end{array}\right)
\end{align}
The quantum and classical fields are defined as usual in the following~\cite{Kamenev}
\begin{align}
\phi^{\rm cl}_j &=\frac{1}{\sqrt 2}(\phi^+_j+\phi^-_j),~~\phi^{\rm q}_j=\frac{1}{\sqrt 2}(\phi^+_j-\phi^-_j)\\
{\phi^{\rm cl}_j}^\ast &=\frac{1}{\sqrt 2}({\phi^+_j}^\ast+{\phi^-_j}^\ast),~~{\phi^{\rm q}_j}^\ast=\frac{1}{\sqrt 2}({\phi^+_j}^\ast-{\phi^-_j}^\ast)\\
\psi_{j1\sigma}&=\frac{1}{\sqrt 2}(\psi^+_{j\sigma}+\psi^-_{j\sigma}),~~\psi_{j2\sigma}=\frac{1}{\sqrt 2}(\psi^+_{j\sigma}-\psi^-_{j\sigma})\\
\bar\psi_{j1\sigma}&= \frac{1}{\sqrt 2}(\bar\psi^+_{j\sigma}-\bar\psi^-_{j\sigma}),~~\bar\psi_{j2\sigma}=\frac{1}{\sqrt 2}(\bar\psi^+_{j\sigma}+\bar\psi^-_{j\sigma})
\end{align}
where $\phi_j^{+(-)}$ and $\psi_{j\sigma}^{+(-)}$ are the bosonic and fermionic fields along the forward (backword) branch of the time contour~\cite{Kamenev}. The fermionic and bosonic propagators are given in Keldysh space as
\begin{align}
G_{Fj} &=\left(\begin{array}{cc}G^R_{Fj}&G^K_{Fj}\\0&G^A_{Fj}\end{array}\right),\\
G_{Bj} &=\left(\begin{array}{cc}G^K_{Bj}&G^R_{Bj}\\G^A_{Bj}&0\end{array}\right).
\end{align}
The retard (advanced) Green function of fermions is given by
\begin{equation}
G^{R(A)}_{Fj}=\frac{1}{\omega-k^2/2m+\mu_j\pm i0^+}
\end{equation}
and the Keldysh Green function
\begin{equation}
G^K_{Fj}=(1-2n_F(\omega))(G^R_{Fj}-G^A_{Fj}).
\end{equation}
The retard (advanced) Green function of bosons in momentum space is given by
\begin{equation}
G^{R(A)}_{Bj}=\frac{1}{\omega-k^2/4m-2\nu+2\mu_j\pm i0^+}
\end{equation}
and the Keldysh Green function is $G^K_{Bj}=(1+2n_B(\omega))(G^R_{Bj}-G^A_{Bj})$, where $n_{F(B)}=(\exp[\beta \omega]\pm1)^{-1}$ is the Fermi (Bose) distribution function.

With a chemical potential bias both fermions and the bosons can be transferred through the QPC. The currents contributed by fermions and bosons are defined as
\begin{align}
I_F(t) &=\frac{1}{2}\left\langle\frac{\partial(N_R^F-N^F_L)}{\partial t}\right\rangle=\frac{1}{2i\hbar}\left\langle[H,N^F_R-N^F_L]\right\rangle,\\
I_B(t) &=\left\langle\frac{\partial(N_R^B-N^B_L)}{\partial t}\right\rangle=\frac{1}{i\hbar}\left\langle[H,N^B_R-N^B_L]\right\rangle.
\end{align}
The factor of two in the above definition is because two fermions are transferred when one boson passes through the QPC. For convenience, we calculate the current in the momentum space. By Fourier transformation, the currents can be expressed as
\begin{widetext}
\bea &&I_F(\Omega)=\frac{1}{A^2}\sum_{n}\int\frac{d\omega}{2\pi}\frac{dk_{L,x}}{2\pi }\frac{dk_{R,x}}{2\pi}\Big\{i\mathcal{T}_F^n(k_{L,x},k_{R,x})\sum_\sigma\hat{ \psi}^\dag_{L\sigma}(\omega, k_{L,x})\hat{\psi}_{R\sigma}(\Omega+\omega+\Delta\mu,k_{R,x})\Big\}+{\rm h.c.},\cr&&
I_B(\Omega)=\frac{1}{A^2}\sum_{n}\int\frac{d\omega}{2\pi}\frac{dk_{L,x}}{2\pi }\frac{dk_{R,x}}{2\pi}\Big\{i\mathcal{T}^n_B(k_{L,x},k_{R,x})\hat{\phi}^\dag_{L}(\omega,k_{L,x})\hat{\phi}_{R}(\Omega+\omega+2\Delta\mu, k_{R,x})\Big\}+{\rm h.c.}.\eea
\end{widetext}

We introducing an time-dependent external source fields $A(t)$ to generate the fermionic and bosonic current. The partition function of the whole system in the momentum space can be written as
\begin{align} \mathcal Z=\frac{1}{\mathcal Z_0}\int D[\bar\psi_{j\sigma},\psi_{j\sigma}, \phi^\ast_j,\phi_j]\exp(iS),
\end{align}
where $S=S_0+S_I+S_T+S_s$ and $S_0$, $S_I$, $S_T$ and $S_s$ correspond to the free, interaction, tunneling and source terms, respectively. They can be expressed in the momentum space as the following
\begin{widetext}
\bea &&S_0=\int \frac{d\omega}{2\pi}\frac{d^3{\bf k}}{(2\pi)^3}\Big\{\sum_\sigma\bar\Psi_{j\sigma}(\omega,{\bf k}) [G_{Fj}]^{-1}\Psi_{j\sigma}(\omega,{\bf k})+\Phi^\ast_j (\omega,{\bf k}) [G_{Bj}]^{-1}\Phi_j(\omega,{\bf k}),\Big\}\cr&& S_I=g\int \frac{d\omega_1}{2\pi}\frac{d\omega_2}{2\pi}\frac{d^3{\bf k}_1}{(2\pi)^3}\frac{d^3{\bf k}_2}{(2\pi)^3}\Big\{\phi^{\rm cl}_j(\omega_1+\omega_2,{\bf k}_1+{\bf k}_2)\big[\bar\psi_{j1\uparrow}(\omega_1,{\bf k}_1)\bar\psi_{j2\downarrow}(\omega_2,{\bf k}_2)+\bar\psi_{j2\uparrow}(\omega_1,{\bf k}_1)\bar\psi_{j1\downarrow}(\omega_2,{\bf k}_2)\big]\cr&&~~~~~~+\phi^{\rm q}_j(\omega_1+\omega_2,{\bf k}_1+{\bf k}_2)\big[\bar\psi_{j1\uparrow}(\omega_1,{\bf k}_1)\bar\psi_{j1\downarrow}(\omega_2,{\bf k}_2)+\bar\psi_{j2\uparrow}(\omega_1,{\bf k}_1)\bar\psi_{j2\downarrow}(\omega_2,{\bf k}_2)\big]+{\rm h.c.} \Big\}\cr&&S_T=J^{\rm q}_F(0)+{J^{\rm q}(0)}_B+{\rm h.c.} ,\cr && S_s=i\int\frac{d\Omega}{2\pi}\Big\{A^{\rm cl}(\Omega)[J_F^{\rm q}(-\Omega)+2J_B^{\rm q}(-\Omega)]+A^{\rm q}(\Omega)[J^{\rm cl}_F(-\Omega)+2J^{\rm cl}_B(-\Omega)]\Big\}+{\rm h.c.}.\eea
\end{widetext}
In the above partition function $A^{\rm cl}=\frac{1}{2}(A^++A^-)$ and $A^{\rm q}=\frac{1}{2}(A^+-A^-)$, where $A^+$ and $A^-$ are the external field along the forward and backward time direction. We have defined
\begin{widetext}
\bea &&J^{\rm cl}_F(\Omega)=\frac{1}{A^2}\sum_n\int\frac{d\omega}{2\pi}\frac{dk_{L,x}}{2\pi }\frac{dk_{R,x}}{2\pi}\cr&&\mathcal{T}_F^n(k_{L,x},k_{R,x})\sum_\sigma\Big(\bar\psi_{L1\sigma}(\omega,k_{L,x})
\psi_{R2\sigma}(\Omega+\omega+\Delta\mu,k_{R,x})+\bar\psi_{L2\sigma}(\omega,k_{L,x})
\psi_{R1\sigma}(\Omega+\omega+\Delta\mu,k_{R,x})\Big),\cr&&J^{\rm q}_F(\Omega)=\frac{1}{A^2}\sum_n\int\frac{d\omega}{2\pi}\frac{dk_{L,x}}{2\pi }\frac{dk_{R,x}}{2\pi}\cr&&\mathcal{T}_F^n(k_{L,x},k_{R,x})\sum_\sigma\Big(\bar\psi_{L1\sigma}(\omega,k_{L,x})
\psi_{R1\sigma}(\Omega+\omega+\Delta\mu,k_{R,x})+\bar\psi_{L2\sigma}(\omega,k_{L,x})
\psi_{R2\sigma}(\Omega+\omega+\Delta\mu,k_{R,x})\Big), \cr &&J^{\rm cl}_B(\Omega)=\frac{1}{A^2}\sum_n\int\frac{d\omega}{2\pi}\frac{dk_{L,x}}{2\pi }\frac{dk_{R,x}}{2\pi}\cr&&\mathcal{T}_B^n(k_{L,x},k_{R,x})\Big(\phi^{\rm cl}_{L}(\omega,k_{L,x})^\ast\phi^{\rm cl}_{R}(\Omega+\omega+2\Delta\mu,k_{R,x})+\phi^{\rm q}_{L}(\omega,k_{L,x})^\ast\phi^{\rm q}_{R}(\Omega+\omega+2\Delta\mu,k_{R,x})\Big),\cr&&J^{\rm q}_B(\Omega)=\frac{1}{A^2}\sum_n\int\frac{d\omega}{2\pi}\frac{dk_{L,x}}{2\pi }\frac{dk_{R,x}}{2\pi}\cr&&\mathcal{T}_B^n(k_{L,x},k_{R,x})\Big(\phi^{\rm cl}_{L}(\omega,k_{L,x})^\ast\phi^{\rm q}_{R}(\Omega+\omega+2\Delta\mu,k_{R,x})+\phi^{\rm q}_{L}(\omega,k_{L,x})^\ast\phi^{\rm cl}_{R}(\Omega+\omega+2\Delta\mu,k_{R,x})\Big).\eea
\end{widetext}

Next, the total current can be calculated by
\begin{equation}
I(t)=\int\frac{d\Omega}{2\pi}e^{i\Omega t}I(\Omega)
\end{equation}
where $I(\Omega)$ is given by
\begin{equation}
I(\Omega)=\left.\frac{i}{2}\frac{\partial \mathcal Z}{\partial A^{\rm q}}\right|_{A^{\rm q}=0}=I_F(\Omega)+I_B(\Omega).
\end{equation}
To order of $\mathcal T_{F(B)}^2$, the currents can be calculated
\begin{align}
I_F(\Omega) &=\frac{1}{2}(\left\langle J^{\rm cl}_{F}(-\Omega)J^{\rm q}_{F}(0)^\ast\right\rangle-\left\langle J^{\rm cl}_{F}(\Omega)^\ast J^{\rm q}_{F}(0)\right\rangle)\\
I_B(\Omega) &=\left\langle J^{\rm cl}_{B}(-\Omega)J^{\rm q}_{B}(0)^\ast\right\rangle-\left\langle J^{\rm cl}_{B}(\Omega)^\ast J^{\rm q}_{B}(0)\right\rangle
\end{align}
To calculate the correlation functions in the above equations we adopt the approximations: $\langle\bar\psi_L\psi_L\bar\psi_R\psi_R\rangle\simeq\langle\bar\psi_L\psi_L\rangle\langle\bar\psi_R\psi_R\rangle$ and $\langle\phi^\ast_L\phi_L\phi^\ast_R\psi_R\rangle\simeq\langle\phi^\ast_L\phi_L\rangle\langle\phi^\ast_R\psi_R\rangle$, where all the two-point correlation functions are the single particle propagators renormalized with self-energy within the NSR scheme. A straightforward calculation yields
\begin{widetext}
\begin{align}
&I_{F}(t)=2\alpha_{F}\epsilon_F^R\int \frac{d\omega}{2\pi}\frac{d\epsilon_L}{\sqrt{\epsilon_L}}\frac{d\epsilon_R}{\sqrt{\epsilon_R}}\cr& \Theta(\epsilon_L-n \omega_y-\bar V_g)\Theta(\epsilon_R-n \omega_y-\bar V_g)A^{F}_L(\omega,\sqrt{2m\epsilon_L}) A^{F}_R(\omega+\Delta\mu,\sqrt{2m\epsilon_R})\big[n_{F}(\omega)-n_{F}(\omega+\Delta\mu)\big],\cr&
I_{B}(t)=2\alpha_{B}\epsilon_F^R\int \frac{d\omega}{2\pi}\frac{d\epsilon_L}{\sqrt{\epsilon_L}}\frac{d\epsilon_R}{\sqrt{\epsilon_R}}\cr& \Theta(\epsilon_L-n \omega_y-\bar V_g)\Theta(\epsilon_R-n \omega_y-\bar V_g)A^{B}_L(\omega,\sqrt{4m\epsilon_L}) A^{B}_R(\omega+2\Delta\mu,\sqrt{4m\epsilon_R})\big[n_{B}(\omega)-n_{B}(\omega+2\Delta\mu)\big].
\end{align}
\end{widetext}

\end{document}